\documentstyle[preprint,aps,epsfig]{revtex}
\begin{document}

\tighten
\draft
\preprint{
\vbox{
\hbox{February 1999}
\hbox{ADP 99-5/T350}
}}

\title{Strange Asymmetries in the Nucleon Sea}

\author{W. Melnitchouk$^{1,2}$ and M. Malheiro$^3$}

\address{$^1$	Institut f\"ur Kernphysik,
		Forschungszentrum J\"ulich,
		D-52425 J\"ulich, Germany}
\address{$^2$	Special Research Centre for the
		Subatomic Structure of Matter,
		University of Adelaide, 5005,
		Australia}
\address{$^3$	Instituto de F\'\i sica,
		Universidade Federal Fluminense,\\
		24210-340, Niter\'oi, Brazil}

\maketitle

\begin{abstract}
Relativistic corrections to the strange axial form factor are
evaluated within a light-cone formalism, taking into account
effects due to the breaking of Lorentz covariance associated
with the use of one-body currents.
Similar corrections are known to also be needed for the magnetic
form factor, and we discuss the importance of these for recent
data on strange electromagnetic form factors from the HAPPEX
Collaboration at Jefferson Lab.
The strange vector form factors, the strange axial charge,
and the asymmetries in strange quark distributions are shown
to be consistently correlated within the chiral cloud model.
\end{abstract}

\section{Introduction}

The sea of the nucleon continues to intrigue nuclear and particle
physicists seeking to understand its structure and dynamical origin.
The recent results from the E866 Collaboration \cite{E866} at Fermilab,
for example, on Drell-Yan production in proton--proton and  
proton--deuteron scattering, from which the $x$-dependence of the
$\bar d/\bar u$ ratio was extracted, present a challenge to models
of the nucleon's structure \cite{MST}.
These data quite unambiguously indicate non-trivial non-perturbative
effects in the proton's sea which give rise to a rather large asymmetry
in the light antiquark sector.

At somewhat lower energies, the HAPPEX Collaboration \cite{HAPPEX}
has recently presented results on the strange electromagnetic form
factors of the proton obtained from parity-violating electron
scattering at Jefferson Lab.
The experiment found
$\left. G_E^S + r G_M^S \right|_{\rm (HAPPEX)} \approx 0.023 \pm 0.048$
at an average $Q^2$ of 0.48 GeV$^2$, with $r \sim 0.4$ for the HAPPEX
kinematics.
This result is consistent with the earlier experiment by the
SAMPLE Collaboration at MIT-Bates \cite{SAMPLE},
$G_{M ({\rm SAMPLE})}^S = + 0.23 \pm 0.44$ at $Q^2 = 0.1$ GeV$^2$.

These experiments are extremely valuable in our quest to arrive at
a consistent picture of the nucleon's substructure.
While valence quark models have provided considerable insight into
the structure of the nucleon's core, describing the dynamics of the
sea of the nucleon is considerably more model-dependent.
Nevertheless, the nucleon sea provides a unique testing ground
for QCD models, since a sea generated purely perturbatively
generally results in vanishing sea quark form factors and
asymmetries.

While constrained by conservation laws requiring equal numbers
of strange and antistrange quarks in the nucleon (which in  
deep-inelastic scattering language corresponds to equal first
moments of the $s$ and $\bar s$ quark distributions, or in
elastic scattering, a zero strange electric form factor at $Q^2=0$),
the distributions of $s$ and $\bar s$ quarks need not be identical
in coordinate or momentum space \cite{SIGNAL,JI}.
Indeed, since perturbative QCD predicts equal $s$ and $\overline s$
distributions, a difference between these would be clear evidence
for non-perturbative effects in the structure of the nucleon.

Many models have been constructed in the literature which attempt
to describe how strangeness arises in the nucleon.
These range from vector meson dominance and quark models, to Skyrme
and NJL types, as well as approaches which try to respect general
properties such as analyticity and chiral symmetry
\cite{JAFFE,PARK,KOEPF,REP,BURK,FORKEL,DEREK,GI,UNIT}.
It is probably fair to say that none of these models is sophisticated
enough or has the sufficient degrees of freedom necessary to provide
a reliable microscopic description of all the strangeness observables.
Nevertheless, in many cases such model studies can offer a glimpse  
of the underlying dynamics of strangeness generation.

A further complication arises with the need to consistently keep
the same model degrees of freedom at different scales.
For example, in deep-inelastic scattering, the natural degrees of
freedom are partons on the light-cone; at low energies one can
obtain reasonable descriptions of observables in terms of effective,
constituent quarks.
Until a rigorous connection is found between these (see however
Ref.\cite{MELOSH}), use of quark-type models will be problematic
if one aims for a unified description of strangeness observables.

A somewhat less ambitious endeavour than calculating structure
functions from first principles is to accept the limitations of
the QCD models, and try to see whether a piece of strangeness
information from one experiment can be used to understand data
from another experiment.
This can also pose some challenges, as the validity of models is
often limited to a specific energy range (for example, below the
chiral symmetry breaking scale, for chiral hadronic models),
forcing one to sometimes extrapolate models to regions where
their reliability could be questioned.

One model which in the past has been applied to the study of both
low energy observables, such as electromagnetic form factors,
magnetic moments, hadron--hadron scattering, etc, as well as
to deep-inelastic structure functions at much higher energies,
is the chiral cloud model \cite{CBM,BROWN,SULL,AWT}.
Since there is {\em a priori} no resolution scale at which chiral
symmetry can be ignored, a cloud of pseudoscalar mesons would be
expected to play some role both at low and high energies, provided
one can isolate the non-perturbative effects from the purely
perturbative ones associated with QCD evolution between different
scales.

One of the main difficulties with the implementation of cloud
models of the nucleon in the past has been how to evaluate
matrix elements of current operators between non-physical
states, such as the virtual mesons or baryons of the cloud.
Some of these can be circumvented by formulating the cloud
on the light-cone (or in the infinite momentum frame).
The light-cone offers many advantages for the description of
hadron and nuclear structure and interactions, as advocated
some 20 years ago by Lepage \& Brodsky \cite{LB} and others.
Interpreting intermediate-state particles as being on their
mass shells, one can avoid introducing {\em ad hoc} off-shell
nucleon form factors or structure functions, and more
consistently parameterise the momentum dependence of vertex 
functions at hadronic vertices.
Furthermore, it is the natural framework for describing partonic
substructure of hadrons and nuclei (see recent work by Miller and
Machleidt \cite{MILLER} in the application of light-cone techniques
to nuclear matter).
Nevertheless, at low energies there are some subtle issues,
such as rotational invariance, which need special care when
dealing with models on the light-cone, and particular care
is paid to these in this paper.

A common assumption in the application of chiral cloud and other
models is the impulse approximation, in which one truncates the
Fock space at some convenient point (usually determined by one's
ability to calculate), and omits contributions from many-body
currents.
It is known, however, that the use of one-body currents alone
for composite systems leads to a violation of Lorentz covariance
\cite{KEISTER,KM}.
In Section II we discuss the consequences of this, and in
particular outline how Lorentz covariance may be restored
using the prescription of Karmanov et al.\cite{KM,KARM,KS}.
More complete accounts can be found, for example,
in Refs.\cite{KREP,SJC}.
In Section III the light-cone chiral cloud model is applied
to the strange axial charge, and important corrections are
found to arise from the Lorentz symmetry breaking effects.
A similar analysis for the strange magnetic form factor was
reported in Ref.\cite{RAP}, where the corrections changed
even its sign, bringing it more into line with the SAMPLE
measurement \cite{SAMPLE}.
The results for the strange electromagnetic form factors from
the chiral cloud model \cite{RAP,MM} are compared with the new
data from the HAPPEX Collaboration \cite{HAPPEX} at Jefferson Lab
in Section IV.
Finally, in Section V we summarise our findings and outline
possible improvements of our analysis in future work.

\section{Relativistic Covariance and the Light-Cone}

As is well known, the light-cone formulation of dynamics has many
advantages over other formulations when dealing with composite
systems.
The more pertinent ones to the current discussion are those
connected with the fact that negative energy contributions to
intermediate states are absent, and particles can be treated
as if they were on-mass-shell.
In practice, this means that matrix elements of hadrons
(and nuclei) can be simply expressed as convolutions of the
constituent particles' matrix elements, and the constituents'
distributions in the hadron.
The issue of off-mass-shellness has plagued many earlier,
instant form calculations which attempted to incorporate
relativistic effects \cite{OFF}.
This is not to say that the light-cone formulation solves
all problems which arise in instant form approaches ---
rather, to some extent one merely reshuffles them according to
what is most convenient for a given application.
Furthermore, if one is to unambiguously correlate information
on strangeness observables from different experiments, one must
utilise the same framework consistently throughout.
Since the light-cone is the appropriate framework for high-energy
deep-inelastic scattering, it would appear natural to also use
the same model on the light-cone to describing observables such
as elastic form factors.

One should point out, however, that the issue of relativistic
covariance is relevant both in light-cone \cite{KM} as well
as instant-form \cite{SJC} approaches, whenever a Fock state
suffers some kind of truncation, which invariably leads to a
violation of Lorentz covariance.
The problem exists because one-body currents, which do not include
interactions, and to which most model calculations are restricted,
do not commute with the interaction-dependent generators of the
Poincar\'e group.
Consequently, an incorrect four-vector structure appears in the
matrix elements of current operators, resulting in the presence
of additional unphysical, or spurious, terms in a form factor
expansion of any electroweak current matrix element.
The spurious form factors would not be present if the Lorentz
symmetry were exact.

In the explicitly covariant formulation of light-cone dynamics
developed by Karmanov et al.\cite{KM,KARM}, a specific method was
proposed for extracting the nucleon's physical form factors,
excluding the spurious contributions.
In this formulation, the nucleon state vector is defined on
a light-cone given by the invariant equation $n \cdot x = 0$,
where $n$ is an arbitrary light-like four vector, $n^2 = 0$.
Since this formulation is covariant, the Lorentz symmetry is
restored, but the matrix elements now depend on the position
of the light-cone plane, $n^\mu$, which in principle no physical
quantity should.
Because of the explicit dependence on the light-cone orientation,
a form factor expansion of the matrix elements of the electroweak
current will now involve three variables, the nucleon ($p^\mu$)
and photon ($q^\mu$) four-momenta and $n^\mu$, rather than the
usual first two.
Therefore in general more structures will appear in the form factor
decomposition, some of whose coefficients (namely, those depending
on $n^\mu$) will be unphysical.
The advantage of this prescription is that these $n^\mu$ dependent
coefficients can then be identified and subtracted from the physical
form factors.

One can also compare the covariant light-cone formulation with
the approach more commonly used in the literature for calculating
light-cone matrix elements, namely using the ``+'' component of
currents, with $n^\mu = (1;0,0,-1)$, so that $t+z=0$ defines
the usual light-cone plane.

In Ref.\cite{KM} the corrections to the electromagnetic vector
form factors of the nucleon were calculated in a quark model.
It was found that while the electric form factor does not suffer
from any contamination from spurious form factors, the magnetic
form factor receives quite large contributions.
Following a similar philosophy, the corrections to the strange
vector form factors of the nucleon were estimated in Ref.\cite{RAP}
within a light-cone chiral cloud model.
For intrinsically small quantities such as those involving
strangeness, any corrections are likely to be relatively
more important, and indeed the strange magnetic form factor
was seen to change sign when the spurious contributions were
subtracted.
In addition to the strange vector matrix elements, those of the
strange axial vector current are also of considerable interest,
as they convey information on the spin distribution of strange
quarks in the nucleon, which has been actively debated since
the discovery of the EMC-spin effect a decade ago \cite{EMC}.
In the following Section we examine the strange axial charge
in the light-cone formulation, and estimate the contamination
to this from the spurious form factors within the chiral cloud
model.

\section{Strange Axial Charge}

Given that one of the main reasons for the focus on the strangeness
content of the nucleon was the distribution of the proton's spin
amongst its constituents, it is clearly important to test whether
previous estimates of strange contributions to the axial charge
are reliable.
In this Section we apply the prescription described above to the
nucleon's strange axial charge.

The strange axial current on the light-cone can be written
covariantly as \cite{KARM}:
\begin{eqnarray}
\label{Jfull}
J_{\mu 5}^S
&=& g_A^S\ \gamma_\mu \gamma_5
 +\ b_1^S\ { \not\!n p_\mu \over n \cdot p} \gamma_5
 +\ \cdots\ ,
\end{eqnarray}
where the ``$\cdots$'' represent terms which do not contribute to
axial matrix elements.
The $b_1^S$ in this decomposition arises precisely because of the
extra $n$ dependence introduced by the light-cone orientation.
In an exact calculation it would be identically zero.
In practice, however, when one uses Lorentz covariance violating
approximations, such as restrictions to one-body currents, this
$n$ dependent form factor can be non-zero.

Taking the forward matrix element of the current $J_{\mu 5}^S$,
one can extract the axial charge $g_A^S$ by using the trace
projection \cite{KARM}:
\begin{eqnarray}
g_A^S &=& {1 \over 4 (n \cdot p)^2} {\rm Tr}
\left[ {\cal O}_\mu
\left( (n \cdot p) p^\mu \not\!n \gamma_5
     - M^2 n^\mu \not\!n \gamma_5
     - (n \cdot p)^2\ \gamma^\mu \gamma_5
\right)
\right],
\end{eqnarray}
where 
${\cal O}_\mu = (\not\!p + M) J_{\mu 5}^S (\not\!p + M)/(4 M^2)$.
Without correcting for the unphysical $b_1^S$ form factor, the
axial charge would be \cite{KARM}:
\begin{eqnarray}
\label{gtilde}
\widetilde g_A^S &=& g_A^S + b_1^S
\ =\ { M^2 \over 2 (n \cdot p)^2 }
{\rm Tr} \left[ {\cal O}_\mu\ n^\mu \not\!n \gamma_5 \right].
\end{eqnarray}

To ascertain the importance of the difference between
$\widetilde g_A^S$ and the corrected $g_A^S$, one can use
a simple chiral cloud model, in which the strangeness in the
nucleon is assumed to reside in the kaon and hyperon components
of the nucleon wave function.
Because of the very different masses and momentum distributions
of the kaon and hyperon, the overall strange and antistrange
distributions will be quite different \cite{SIGNAL}.
In particular, in the valence approximation for the cloud,
the $\bar s$ distribution is expected to be zero, since it
resides entirely in the scalar kaon.

In the chiral cloud model the nucleon couples to a pseudoscalar
kaon ($K$) and a spin-1/2 hyperon ($Y$) via a pseudoscalar
$i\gamma_5$ interaction (the same results are also obtained
with a pseudovector coupling).
Extension of this analysis to spin-3/2 hyperons or strange
vector mesons is straightforward, although beyond the scope
of the present discussion (see below).
Because the kaon has spin 0, the axial form factor receives
contributions only from the $\gamma^* \Lambda$ coupling,
which can be written:
\begin{eqnarray}
\label{gas}
g_A^S &=& { g_{K N Y}^2 \over 16 \pi^3 }
\int { dy\ d^2{\bf k}_T \over y^2 (1-y) }
{ {\cal F}^2 \over ({\cal M}^2 - M^2)^2 }
\left( 1 - { M_Y \over y M } \right)
\left( k_T^2 + M_Y (M_Y - y M) \right),
\end{eqnarray}
where ${\cal M}^2 = (k_T^2 + M_Y^2)/y + (k_T^2 + m_K^2)/(1-y)$
is the invariant mass of the intermediate state, and ${\cal F}$
parameterises the hyperon-meson-nucleon vertex.

The momentum dependence of the vertex function ${\cal F}$ can be
calculated within the same model by dressing and renormalising
the bare $KNY$ vertex by $K$ loops.
However, since a detailed model description of the hadronic vertex
is not the purpose of this paper, we shall instead follow the more
phenomenological approach and parameterise the $KNY$ vertex by a
simple function, such as a monopole,
${\cal F} = (\Lambda^2 + M^2) / (\Lambda^2 + {\cal M}^2)$.
We shall comment on the dependence of the strangeness distribution
on the shape of the form factor later.

For the uncorrected strange axial charge, from Eq.(\ref{gtilde})
we have:
\begin{eqnarray}
\label{gastilde}
\widetilde g_A^S &=& { g_{K N Y}^2 \over 16 \pi^3 }
\int { dy\ d^2{\bf k}_T \over y^2 (1-y) }
{ {\cal F}^2 \over ({\cal M}^2 - M^2)^2 }
\left( -k_T^2 + (M_Y - y M)^2 \right),
\end{eqnarray}
which agrees with the expressions obtained in Ref.\cite{MM}.
The results for the strange axial charge $g_A^S$ are shown in Fig.1
as a function of the cut-off mass $\Lambda$.
In practice, the $K\Lambda$ configuration turns out to give by far
the dominant contribution to $g_A^S$ if standard coupling constants
\cite{HOLZ} and form factor cut-offs are used.
Also shown in Fig.1 is the uncorrected charge $\widetilde g_A^S$.
The $n$ dependent form factor turns out to be rather large,
and contaminates the ``true'' $g_A^S$ to such an extent so
as to produce the rather small $\widetilde g_A^S$ value
observed.
The only empirical information available on the strange axial
charge comes from the Brookhaven 734 experiment \cite{BNL} on
elastic $\nu p$ and $\bar \nu p$ scattering \cite{KAPLAN}.
Unfortunately, the value of $g_A^S$ extracted from this experiment
was found to be strongly correlated with the value of the cut-off
mass, $M_A$, in the dipole axial vector form factor parameterisation.
Varying $M_A$ between $1.086 \pm 0.015$ GeV and $1.012 \pm 0.032$ GeV,
one can obtain anything between $g_A^S = 0$ and $-0.21 \pm 0.10$
\cite{GARVEY}, as indicated by the shaded region in Fig.1.

One can also compare the strange axial charge with the first moment,
$\Delta s$, of the polarised strange quark distribution measured in
deep-inelastic scattering \cite{DIS}.
A recent world-averaged value extracted in the $\overline{\rm MS}$
scheme at a scale of 10 GeV$^2$ is $\Delta s = -0.10 \pm 0.04$
\cite{DIS}, as indicated by the two long-dashed horizontal lines
in Fig.1.
Note that in the chiral cloud model the distribution $\Delta s$
is given by a convolution of the $y$-integrand in Eqs.(\ref{gas})
and (\ref{gastilde}) with the polarised strange distribution in
the hyperon \cite{MM}.
Since the latter is not a $\delta$-function, but has a non-trivial
$x$-dependence, the resulting convolution would be expected to be
smaller than the strange axial charge in the model.
The experimental value of $\Delta s$ is nonetheless consistent
with the calculated $g_A^S$ if a soft $KNY$ form factor is used.

To constrain the size of the $KNY$ form factor, which is essentially
the only parameter in the chiral cloud model, one can compare the
model predictions with the measured unpolarised $s$--$\bar s$
asymmetry.
The possible differences between the $s$ and $\bar s$ quark
distributions in the nucleon were investigated by the CCFR
Collaboration via charm production in $\nu$ and $\bar \nu$
deep-inelastic scattering \cite{CCFR}.
Such differences were first predicted in the meson cloud
framework more than 10 years ago by Signal and Thomas
\cite{SIGNAL}.
The $x$-dependence of the calculated $s$--$\bar s$ distribution
is shown in Fig.2 for a form factor cut-off mass of $\Lambda=1$ GeV
(which gives an average multiplicity of kaons in the nucleon of
$\approx 6\%$).
The shaded region represents the data from Ref.\cite{CCFR}.
Also shown for comparison is the result with a $t$-dependent
monopole form factor, as used in earlier analyses,
${\cal F} = (\Lambda - m_K^2)/(\Lambda - t)$, where
$t = (p_N - p_Y)^2$.
Notice that the final shape and sign of $s$--$\bar s$ are quite
sensitive to the shape of the $KNY$ form factor.
On the other hand, it is known that form factors which depend
solely on the $t$ variable violate momentum conservation when
one considers scattering from both the meson and hyperon components
of the nucleon \cite{MT}.
More precise measurement of the strange asymmetry would be a
valuable test of the dynamics of the $KNY$ interaction.

One can also compare the predictions of the model with the
absolute values of the extracted $s$ and $\bar s$ distributions,
as done in Ref.\cite{JI} for example.
As in Fig.2, one finds that for a hard hadronic form factor
the meson cloud contributions overestimate the data, especially
at large $x$ \cite{JI}.
The problem with comparing to the total $s$ and $\bar s$ distributions,
however, as distinct from their difference, is that the total
distributions contain singlet contributions in addition to the
non-singlet.
Modeling the former in general requires the (symmetric) perturbative
sea arising from $g \rightarrow s \bar s$, as well as additional input
for the structure of the bare nucleon distributions, uncertainties
in which consequently make any real predictions of the model more
elusive.
For this reason a comparison with the non-singlet difference
$s-\bar s$, in which the perturbative contributions cancel,
is more meaningful for the meson cloud model.

Finally, before ending this Section, we should note several
concerns which have been raised in the literature regarding the
implementation of loops in chiral models of the nucleon.
In particular, it has been pointed out that truncations of the
Fock state which stop at the one-loop order violate, in addition
to the Lorentz covariance discussed above, also unitarity 
\cite{UNIT}.
While this is true in principle, the region where rescattering
should become an issue is above the production threshold,
which in practice is at rather high momenta compared with
those most relevant to the current process.
Furthermore, the chiral cloud model discussed here rests on a
perturbative treatment of the effective hadronic Lagrangian,
so that provided the form factors used at the hadronic vertices
are not very hard, one would expect a one-loop calculation for
the most part to give the dominant contribution.
If two loop contributions were found to be large compared with
the leading ones, the perturbative formulation of the chiral
cloud itself would need to be reconsidered.
Recent work \cite{FIOLHAIS} based on coherent states techniques
which include effects of higher-order, multi-pion Fock states for
models such as the cloudy bag \cite{CBM} indicates that for
relatively small meson densities, a one-loop, perturbative
treatment comes very close to the exact result.
The conclusion is therefore that so long as the hadronic vertices
are relatively soft, with $\Lambda \sim 1$ GeV, the one-loop result
should give a reasonable estimate of cloud effects.

Concerns have also been raised about the omission of contributions
from higher-mass intermediate states in the meson--baryon fluctuations
\cite{KSTAR}.
While the effects of heavier baryons such as the $\Sigma^*$
have been shown to be negligible \cite{MM}, it has been argued
that strange vector meson contributions are of the same order
of magnitude as the $K$.
In the analysis of Ref.\cite{KSTAR} a rather hard $K^* N\Lambda$
form factor was used, however, with a cut-off mass in the monopole
parametrisation of $\Lambda \sim 2.2$ GeV.
This is to be compared with $\Lambda \sim 1.2$ GeV for the
$KN\Lambda$ vertex \cite{KSTAR}.
This relatively large value for the $K^* N\Lambda$ cut-off
was taken from the hyperon--nucleon scattering analysis of
Ref.\cite{HOLZ}, although more recent work \cite{YN,NIJM}
suggests that a value for {\em both} the $K^* N \Lambda$ and
$KN\Lambda$ form factor cut-offs of $\sim 1$~GeV is more
appropriate, 
Such a smaller value would significantly reduce the $K^*$
contribution.
A re-evaluation of the strange vector meson effects with softer
form factors would therefore be very useful before definitive
conclusions about the reliability of lowest-order one-loop
calculations can be made.

\section{Strange Electromagnetic Form Factors}

As well as understanding polarised strangeness in the nucleon,
there has also been considerable effort directed at measuring
matrix elements of the electromagnetic vector currents.
The first experimental result on the strange magnetic form
factor of the proton was obtained by the SAMPLE Collaboration
\cite{SAMPLE} at MIT-Bates in 1997 in parity-violating electron
scattering at backward angles, at $Q^2=0.1$ GeV$^2$.
While plagued with large errors, the data did seem to favour
a relatively small, and possibly positive, value of the strange
magnetic moment.
More recently, the HAPPEX Collaboration at Jefferson Lab
\cite{HAPPEX} performed a similar experiment, although at
forward angles, measuring the left-right asymmetry $A$ at
$Q^2=0.48$ GeV$^2$, where:
\begin{eqnarray}
A &=& { \sigma_R - \sigma_L \over \sigma_R + \sigma_L }
\  =\
\left( { - G_F \over \pi \alpha_{em} \sqrt{2} } \right)
{ 1 \over \varepsilon\ G_E^2 + \tau\ G_M^2 }		\nonumber\\
  & & \times
\left( \varepsilon\ G_E\ G_E^{(Z)}
     + \tau\     G_M\ G_M^{(Z)}
     - {1\over 2} (1 - 4 \sin^2\theta_W)\ \varepsilon'\ G_M\ G_A^{(Z)}
\right),
\end{eqnarray}
with
$\varepsilon = \left( 1 + 2 (1+\tau) \tan^2(\theta/2) \right)^{-1}$,
$\tau = Q^2/4 M^2$,
and $\varepsilon' = \sqrt{ \tau (1+\tau) (1-\varepsilon^2) }$
(the $Q^2$ dependence in all form factors is implicit).

Using isospin symmetry, one can relate the electric and magnetic
form factors for photon and $Z$-boson exchange via:
\begin{eqnarray}
G_{E,M}^{(Z)} &=& {1 \over 4} G_{E,M}^{(I=1)}
	       -  \sin^2\theta_W\ G_{E,M}
	       -  {1 \over 4} G_{E,M}^S,
\end{eqnarray}
where $G_{E,M}^{(I=1)}$ is the isovector form factor (difference
between the proton and neutron).
For the $G_{E,M}$ form factors we use the parameterisation from
Ref.\cite{MMD}.
The axial form factor for $Z$-boson exchange is given by
$G_A^{(Z)} = -{1\over 2} (1+R_A) G_A + {1\over 2} G_A^S$,
where $R_A$ is an axial radiative correction, and the axial
form factors are known phenomenologically \cite{REP}.

In Fig.3 we plot the relative difference between the measured
asymmetry, $A$, and that which would be expected for zero strangeness,
$A_0$ --- namely, $(A-A_0)/A$.
The solid curve corresponds to the light-cone chiral cloud model
with a cut-off mass $\Lambda=1$ GeV for the kaon--hyperon vertex.
{}From the measured HAPPEX asymmetry, the combination $G_E^S + r G_M^S$
was also extracted at an average $Q^2$ of 0.48 GeV$^2$,
where $r = (\tau/\varepsilon) G_M/G_E \approx 0.4$
for the HAPPEX kinematics.
This is shown in Fig.4, compared with the chiral cloud prediction
with $\Lambda=1$ GeV.
Both the magnetic (dotted) and electric (dashed) contributions
are separately positive, resulting in a small and positive value,
consistent with the experiment.
Note that exactly the same parameters were used in Figs.3 and 4
as in the fit in Ref.\cite{RAP} to the SAMPLE data on $G_M^S$.
Therefore the two form factor measurements, as well as the strange
axial charge and the strange--antistrange asymmetry, seem to be
consistently correlated within the chiral cloud model with soft
form factors.

\section{Conclusion}

In this note we have pointed out the existence of corrections to
the strange axial charge of the nucleon which arise in light-cone
models based on the impulse approximation, or one-body operators,
in which Lorentz covariance is not preserved.
In the chiral cloud model, where the strangeness content of the
nucleon is localised to the kaon--hyperon components of the
nucleon wave function, these corrections are an order of magnitude
larger than the uncorrected, Lorentz-violating results, and compatible
with the sign and magnitude of the empirical $g_A^S$.

With the same model parameters, namely a soft kaon--hyperon--nucleon
form factor (with a kaon probability in the nucleon of $\alt 6\%$),
one also has good agreement with the strange electromagnetic
form factors measured in recent experiments at low $Q^2$ at
MIT-Bates \cite{SAMPLE} and Jefferson Lab \cite{HAPPEX}.
The results are also compatible with data on the strange--antistrange
asymmetry from the CCFR experiments \cite{CCFR}.

One should of course mention some of the shortcomings of the
simple one-loop meson cloud model treatment, which may qualify
some of the quantitative predictions of the model.
One of these is the problem of gauge invariance, which in earlier,
instant-form approaches has been partially circumvented with
the inclusion of contact, or so-called seagull, terms \cite{SEAGULL}.
Unfortunately, these are not unique \cite{WANG,HABERZ}, and to date
one does not have control over the size of these contributions.
Other potential contributions may arise from heavier meson
Fock states (such as $K^*$) or multi-meson configurations.
These will be more quantitatively analysed in future work,
but our previous experience suggests that their effects are
unlikely to be dramatic in a perturbative treatment.

More theoretical work is obviously needed for a deeper understanding
of the dynamics of strangeness generation in the nucleon.
What seems to be becoming clearer, however, from the accumulating
empirical evidence is that the importance of non-perturbative
strangeness in the nucleon is likely to be relatively minimal.
Future data from Jefferson Lab on the strange electromagnetic form
factors, $G_{E,M}^S$, over a range of $Q^2$ should help to clarify
this further.

\acknowledgements

We would like to thank C. Boros, M.J. Ramsey-Musolf, F.M. Steffens
and A.W. Thomas for helpful comments and discussions, and
O. Melnitchouk for a careful reading of the manuscript.
M.M. is	partially supported by CNPq of Brazil.


\begin{figure}
\label{fig1}
\epsfig{figure=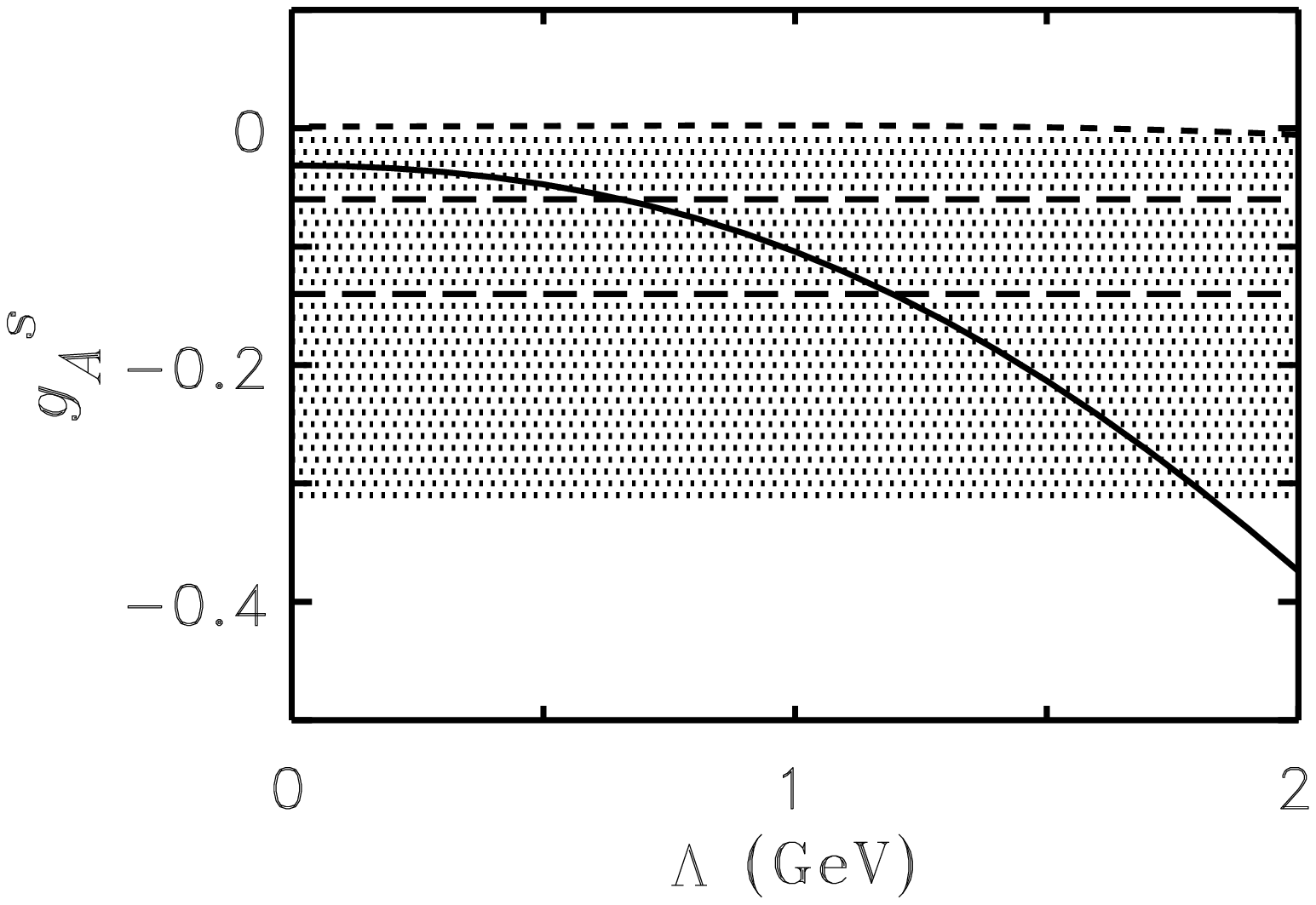,height=9cm}
\caption{Strange axial charge of the proton as a function of
	the hadronic vertex function cut-off mass, $\Lambda$.
	The solid line is the full result from Eq.(\protect\ref{gas}),
	while the short-dashed is the uncorrected result from
	Eq.(\protect\ref{gastilde}).
	The shaded area represents the range for $g_A^S$ found
	in $\nu p$ and $\bar \nu p$ elastic scattering
	\protect\cite{GARVEY}, and the two long-dashed lines
	are the limits on $\Delta s$ from deep-inelastic
	scattering \protect\cite{DIS}.}
\end{figure}

\begin{figure}
\label{fig2}
\epsfig{figure=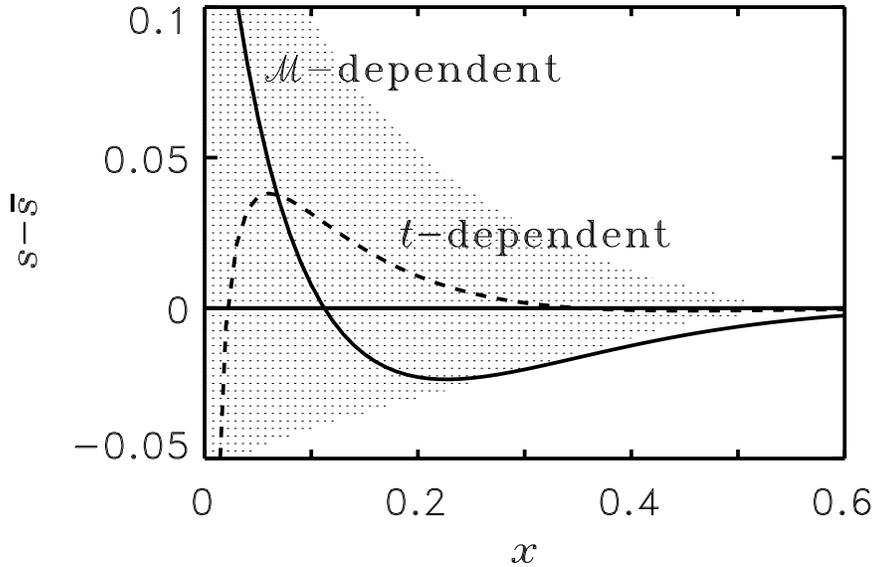,height=9cm}
\caption{Strange -- antistrange quark difference in the nucleon,
	with ${\cal M}$-dependent (solid) and $t$-dependent (dashed)
	monopole form factors, each with a $\Lambda = 1$ GeV momentum
	cut-off (giving a normalisation of
	$\langle n \rangle_{KY} \approx 6\%$).}
\end{figure}

\begin{figure}
\label{fig3}
\epsfig{figure=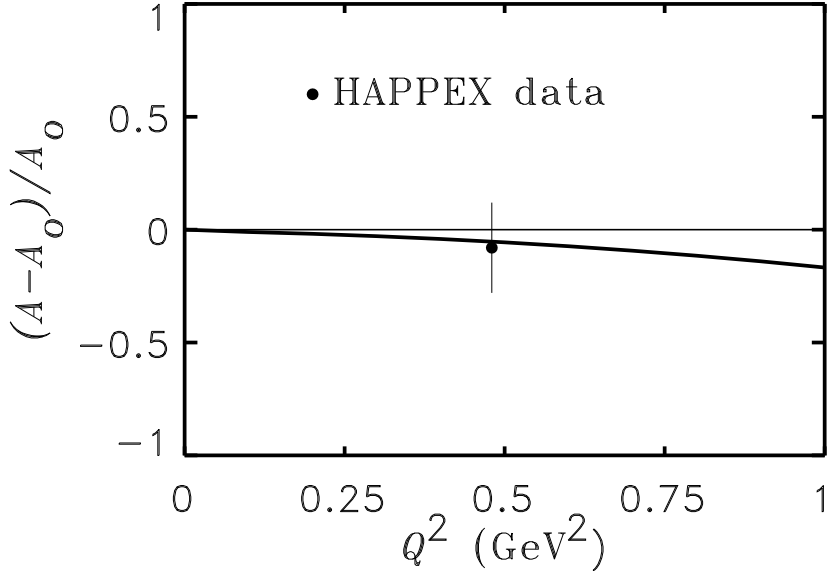,height=9cm}
\caption{Relative difference between the measured \protect\cite{HAPPEX}
	left-right asymmetry, $A$, and that expected for zero strangeness,
	$A_0$.
	The solid curve is evaluated with the chiral cloud values
	for $G_{E,M}^S$ with $\Lambda=1$ GeV.}
\end{figure}

\begin{figure}
\label{fig4}
\epsfig{figure=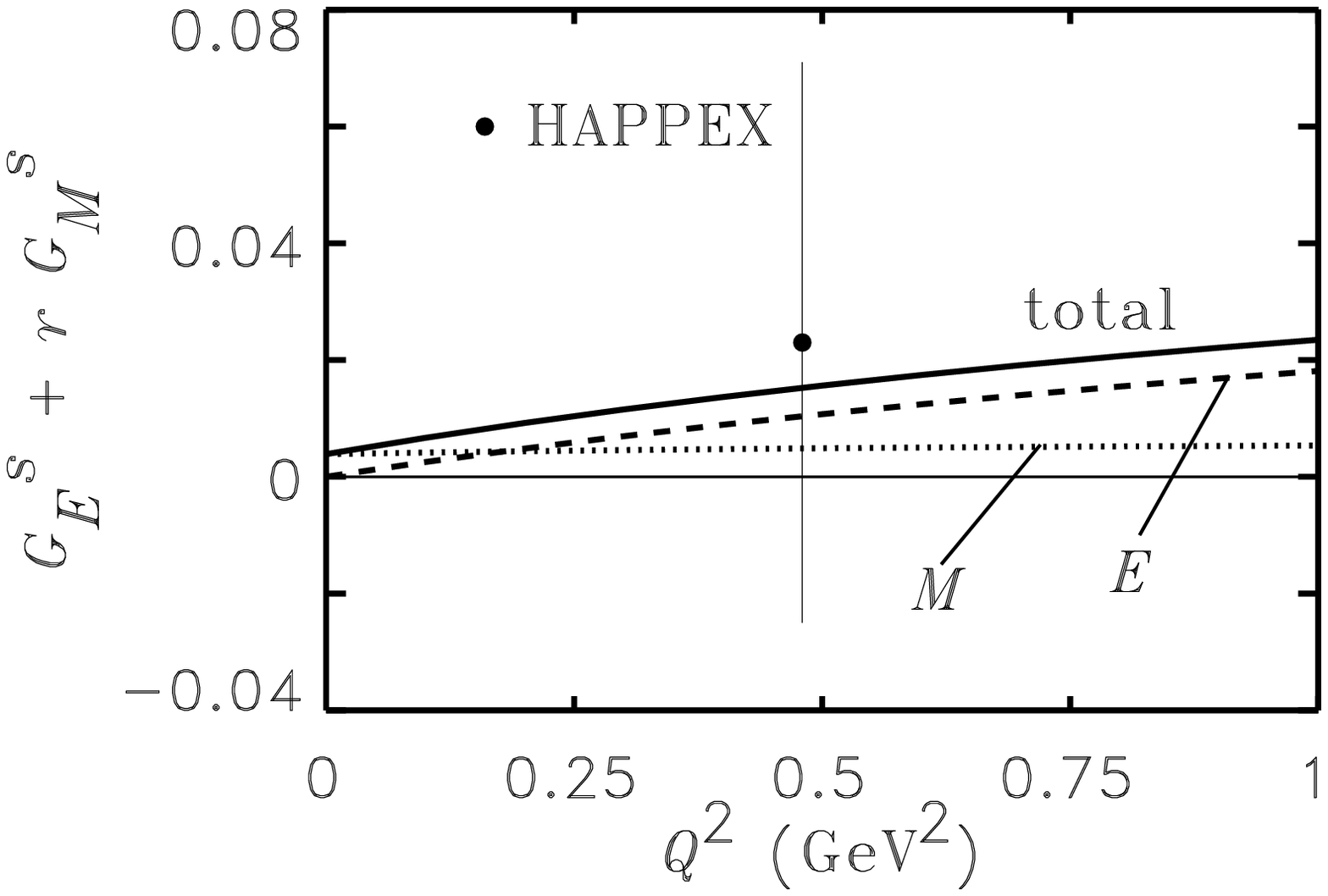,height=9cm}
\caption{Strange electric and magnetic form factor combination
	as extracted from the HAPPEX data \protect\cite{HAPPEX}.
	The curves are for the chiral cloud model with a form
	factor cut-off of $\Lambda=1$ GeV.}
\end{figure}


\references

\bibitem{E866}
E866/NuSea Collaboration, E.A. Hawker et al.,
Phys. Rev. Lett. 80 (1998) 3715;
%
J.C. Peng et al.,
Phys.Rev. D 58 (1998) 092004.

\bibitem{MST}
W. Melnitchouk, J. Speth and A.W. Thomas,
Phys. Rev. D 59 (1999) 014033.

\bibitem{HAPPEX}
HAPPEX Collaboration, K.A. Aniol et al.,
nucl-ex/9810012, to appear in Phys. ReV. Lett.

\bibitem{SAMPLE}
SAMPLE Collaboration, B. Mueller et al.,
Phys. Rev. Lett. 78 (1997) 3824.

\bibitem{SIGNAL}
A.I. Signal and A.W. Thomas,
Phys. Lett. B 191 (1987) 206.

\bibitem{JI}
X. Ji and J. Tang,
Phys. Lett. B 362 (1995) 182.

\bibitem{JAFFE}
R.L. Jaffe,
Phys. Lett. B 229 (1989) 275.

\bibitem{PARK}
N.W. Park, J. Schechter and H. Weigel,
Phys. Rev. D 43 (1991) 869.

\bibitem{KOEPF}
W. Koepf, E.M. Henley, and S.J. Pollock,
Phys. Lett. B 288 (1992) 11.

\bibitem{REP}
M.J. Musolf, T.W. Donnelly, J. Dubach, S.J. Pollock, S. Kowalski
and E.J. Beise,
Phys. Rep. 239 (1994) 1.

\bibitem{BURK}
M.J. Musolf and M. Burkardt,
Z. Phys. C 61 (1994) 433.

\bibitem{FORKEL}
H. Forkel, M. Nielsen, X. Jin and T.D. Cohen,
Phys. Rev. C 50 (1994) 3108.

\bibitem{DEREK}
D.B. Leinweber,
Phys. Rev. D 53 (1996) 5115.

\bibitem{GI}
P. Geiger and N. Isgur,
Phys. Rev. D 55 (1997) 299.

\bibitem{UNIT}
M.J. Ramsey-Musolf and H.W. Hammer,
Phys. Rev. Lett. 80 (1998) 2539;
%
M.J. Musolf, H.W. Hammer and D. Drechsel,
Phys. Rev. D 55 (1997) 2741.

\bibitem{MELOSH}
H.J. Melosh,
Phys. Rev. D 9 (1974) 1095.

\bibitem{CBM}
S. Theberge, G.A. Miller and A.W. Thomas,
Can. J. Phys. 60 (1982) 59;
%
A.W. Thomas,
Adv. Nucl. Phys. 13 (1984) 1.

\bibitem{BROWN}
G.E. Brown, A.D. Jackson, M. Rho and V. Vento,
Phys. Lett. 140 B (1984) 285;
%
V. Vento, M. Rho, E.M. Nyman, J.H. Jun and G.E. Brown,
Nucl. Phys. A345 (1980) 413.

\bibitem{SULL}
J.D. Sullivan,
Phys. Rev. D 5 (1972) 1732.

\bibitem{AWT}
A.W. Thomas,
Phys. Lett. 126 B (1983) 97.

\bibitem{LB}
G.P. Lepage and S.J. Brodsky,
Phys. Rev. D 22 (1980) 2157.

\bibitem{MILLER}
G.A. Miller and R. Machleidt,
nucl-th/9811050.

\bibitem{KEISTER}
B.D. Keister,
Phys. Rev. D 49 (1994) 1500;
%
Phys. Rev. C 55 (1997) 2171.

\bibitem{KM}
V.A. Karmanov and J.-F. Mathiot,
Nucl. Phys. A602 (1996) 388.

\bibitem{KARM}
V.A. Karmanov,
Nucl. Phys. A644 (1998) 165.

\bibitem{KS}
V.A. Karmanov and A.V. Smirnov,
Nucl. Phys. A546 (1992) 691.

\bibitem{KREP}
J. Carbonell, B. Desplanques, V.A. Karmanov and J.F. Mathiot,
Phys. Rep. 300 (1998) 215.

\bibitem{SJC}
A. Szczepaniak, C. Ji and S.R. Cotanch,
Phys. Rev. D 52 (1995) 2738; 5284.

\bibitem{RAP}
M. Malheiro and W. Melnitchouk,
Phys. Rev. C 56 (1997) 2373.

\bibitem{MM}
W. Melnitchouk and M. Malheiro,
Phys. Rev. C 55 (1997) 431.

\bibitem{OFF}
W. Melnitchouk, A.W. Schreiber and A.W. Thomas,
Phys. Rev. D 49 (1994) 1183.

\bibitem{EMC}
European Muon Collaboration, J.Ashman et al.,
Nucl. Phys. B328 (1989) 1. 

\bibitem{HOLZ}
B. Holzenkamp, K. Holinde and J. Speth,
Nucl. Phys. A500 (1989) 485.

\bibitem{BNL}
L.A. Ahrens et al.,
Phys. Rev. D 35 (1987) 785.

\bibitem{KAPLAN}
D.B. Kaplan and A. Manohar,
Nucl. Phys. B310 (1988) 527.

\bibitem{GARVEY}
G.T. Garvey, W.C. Louis and D.H. White,
Phys. Rev. C 48 (1993) 761.

\bibitem{DIS}
E143 Collaboration, K. Abe et al,
Phys. Lett. B 364 (1995) 61;
%
E142 Collaboration, P.L. Anthony et al.,
Phys. Rev. D 54 (1996) 6620;
%
Spin Muon Collaboration, D. Adams et al.,
Phys. Rev. D 56 (1997) 5330.

\bibitem{CCFR}  
A.O. Bazarko et al.,
Z. Phys. C 65 (1995) 189.

\bibitem{MT}
W. Melnitchouk and A.W. Thomas,
Phys. Rev. D 47 (1993) 3794;
%
A.W. Thomas and W. Melnitchouk,
New Frontiers in Nuclear Physics (World Scientific, 1993), pp.41-106.

\bibitem{FIOLHAIS}
M. Fiolhais,
private communication.

\bibitem{KSTAR}
L.L. Barz, H. Forkel, H.-W. Hammer, F.S. Navarra, M. Nielsen
and M.J. Ramsay-Musolf,
Nucl. Phys. A640 (1998) 259.

\bibitem{YN}
J. Haidenbauer, W. Melnitchouk and J. Speth,
Proceedings of Sendai International Workshop on the Spectroscopy
of Hypernuclei, edited by H. Tamura et al.
(Tohoku University Press, Sendai 1998), p.27,
nucl-th/9805014.

\bibitem{NIJM}
P.M.M. Maessen, T.A. Rijken and J.J. de Swart,
Phys. Rev. C 40 (1989) 2226;
%
Th.A. Rijken, V.G.J. Stoks and Y. Yamamoto,
U. Adelaide report ADP-98-37-T310, nucl-th/9807082.

\bibitem{MMD}
P. Mergell, U.-G. Mei\ss ner and D. Drechsel,
Nucl. Phys. A596 (1996) 367.

\bibitem{SEAGULL}
K. Ohta,
Phys. Rev. C 40 (1989) 1335; 
%
F. Gross and D.O. Riska,
Phys. Rev. C 36 (1987) 9128.

\bibitem{WANG}
S. Wang and M.K. Banerjee,
Phys. Rev. C 54 (1996) 2883.

\bibitem{HABERZ}
H. Haberzettl,
Phys. Rev. C 56 (1997) 2041;
%
H. Haberzettl, C. Bennhold, T. Mart and T. Feuster,
Phys. Rev. C 58 (1998) R40.
\end{document}